\documentclass{article}

\usepackage[english]{babel}

\usepackage[letterpaper,top=2cm,bottom=2cm,left=3cm,right=3cm,marginparwidth=1.75cm]{geometry}

\usepackage{amsmath}
\usepackage{graphicx}
\usepackage[colorlinks=true, allcolors=blue]{hyperref}
 \usepackage{multirow}

\title{JPEG Steganalysis Based on Steganographic Feature Enhancement and Graph Attention Learning}
\author{Qiyun Liu, Zhiguang Yang and Hanzhou Wu\footnote{Corresponding author, Email: h.wu.phd@ieee.org}}

\begin{document}

\maketitle

\begin{abstract}
The purpose of image steganalysis is to determine whether the carrier image contains hidden information or not. Since JEPG is the most commonly used image format over social networks, steganalysis in JPEG images is also the most urgently needed to be explored. However, in order to detect whether secret information is hidden within JEPG images, the majority of existing algorithms are designed in conjunction with the popular computer vision related networks, without considering the key characteristics appeared in image steganalysis. It is crucial that the steganographic signal, as an extremely weak signal, can be enhanced during its representation learning process. Motivated by this insight, in this paper, we introduce a novel representation learning algorithm for JPEG steganalysis that is mainly consisting of a graph attention learning module and a feature enhancement module. The graph attention learning module is designed to avoid global feature loss caused by the local feature learning of convolutional neural network and reliance on depth stacking to extend the perceptual domain. The feature enhancement module is applied to prevent the stacking of convolutional layers from weakening the steganographic information. In addition, pretraining as a way to initialize the network weights with a large-scale dataset is utilized to enhance the ability of the network to extract discriminative features. We advocate pretraining with ALASKA2 for the model trained with BOSSBase+BOWS2. The experimental results indicate that the proposed algorithm outperforms previous arts in terms of detection accuracy, which has verified the superiority and applicability of the proposed work.
\end{abstract}

\section{Introduction}
As an important branch of information hiding \cite{IH:ProcIEEE:1999}, steganography is referred to as the art of embedding secret information into a cover signal such as digital image and video without significantly distorting the cover signal. Steganography can be used to effectively solve security problems in digital communication due to its concealment. Generally, any type of digital signal can be used as the cover for steganography. However, from the viewpoint of applications, digital image is the most popular cover for steganography due to its ease of editing and large redundant space for accommodating secret information. As a result, the majority of existing steganographic methods are designed for digital images \cite{Wu:MTAP:2015}. There are different formats for digital images, among which JPEG is the most popular image format over social networks since it provides high visual quality while keeping the storage size of the image low. By applying JPEG images for steganography, the presence of steganographic communication can be easily concealed by the large amount of normal social activities, promoting JPEG steganography to be a hot research spot.

Steganalysis, as a counter technology to steganography, has also been substantially studied with the aim of revealing the existence of hidden information in digital images. Most of the existing steganalysis methods model the detection of hidden information in digital images as a binary classification problem, i.e., to determine whether a given image contains secret information or not. The rapid development of JPEG steganography has motivated JPEG steganalysis to be the most urgently needed to be explored. Early JPEG steganalysis methods \cite{Holub:TIFS:2015, Song:IH:2015, Holub:EI:2015} follow the traditional machine learning framework, which requires manually crafted statistical features and may apply the ensemble strategy. With the popularity of deep convolutional neural networks (CNNs), increasing works \cite{Xu:SPL:2016, Xu:IH:2016, Xu:IH:2017, Su:MTAP:2021} aim at moving the success brought by CNNs in computer vision to image steganalysis in recent years. Generally, there are three stages for CNN based image steganalysis \cite{Liu:MMSP:2022}. First, a certain number of noise residuals are extracted from a given image. Then, a set of high-dimensional feature vectors are determined from these noise residuals. Finally, with a binary classifier, the image can be classified as \emph{cover} or \emph{stego} (which contains secret information). Thanks to the powerful learning ability of CNNs, the aforementioned three stages can be implemented by an end-to-end fashion. Regarding the extraction of image noise residuals, many CNN-based image steganalyzers use a priori information to guide model convergence \cite{Su:MTAP:2021, Liu:Math:2021, Ye:TIFS:2017, Huang:IH:2019}. For example, Huang \emph{et al.} \cite{Huang:IH:2019} recently introduce a steganalysis network that applies selection channels to the spatial domain of JPEG images. Boroumand \emph{et al.} \cite{Boroumand:TIFS:2019} propose SRNet to reduce the reliance of the network on a priori information. This architecture uses a combination of a significantly extended front-end with multiple residual connections to extract noise residuals instead of traditional manual design, resulting in excellent detection performance.

In exploring the applicability of deep learning to steganalysis, it is necessary to take into account steganographic principles. Spatial steganography directly alters the image pixels in the original spatial domain. During data embedding, the pixels to be embedded are varying from images. Neural networks that are powerful in mining local patterns risk remembering specific embedded patterns, which may eventually impair the generalization ability of the trained models. In contrast, JPEG steganography tends to embed secret information in the DCT domain by modifying the quantized DCT coefficients. When transformed back to the spatial domain, the modification of the DCT coefficients extends to all spatial pixels in the corresponding $8\times 8$ block. By learning statistical inconsistencies between blocks in the spatial domain, JPEG steganography is likely to be detected easier than spatial steganography. Motivated by the fact that graph neural networks (GNNs) \cite{Zhou:AIOpen:2020} have powerful ability to model statistical dependencies between objects, we would like to further explore the applicability of graph representation learning to JPEG steganalysis. GNNs have flourished recently because of their superiority in processing non-Euclidean data, and increasing researchers have started to integrate them with visual tasks \cite{Han:NIPS:2022}. Because of its ability to model an image as a complete graph for global analysis, graph representation learning can better grasp the overall statistical characteristics and relevance. Our previous work already demonstrate that graph representation learning can be combined with spatial image steganalysis \cite{Liu:MMSP:2022}, so further exploring its suitability for JPEG steganalysis is urgently to be investigated.

Previous CNN-based works have made a series of explorations in network structures applicable to JPEG steganalysis. While these studies incorporate the existing design concepts of deep learning, they lack attention to the key factors involved in image steganalysis. The problem of feature disappearance during network deepening has been neglected. In image steganalysis, the stego features to be recognized are extremely weak, while convolution and pooling also suppress weak signals. The weak stego features are weakened layer by layer by successive stacks of convolution and pooling in the network, and some may even disappear. From another perspective, CNNs are used to expand the perceptual field of local perception of their convolutional kernels by their depth. In this process, some global information is inevitably lost, which can also be understood as the interrelationship between all the information of the entire image. We believe that the above two factors are the main reasons for the limited performance of existing steganalysis models.

Based on the aforementioned analysis, we believe that two key points need to be considered when designing a network structure that can make better use of stego features. The first point is how to prevent weak stego signals from being suppressed by the network itself, and the second one is how to exploit the global features of stego signals. To this end, in this paper, we develop a new deep learning framework for JPEG steganalysis that combines the advantages of CNN and GNN. For the first point, we apply a feature enhancement block as the basic unit, which removes the pooling layer and reduces the stride of the convolution to 1 to better enhance the weak stego signal. To deal with the problem of how to utilize global features, we feed the image into a graph attention network to learn the pixel-level attention coefficients. Such design can further extract the residual feature vectors effectively from the global perspective and learn the stego features effectively from the global perspective. In addition, pre-training as a way to initialize the network weights with a large-scale dataset can speed up model convergence and enhance the ability of the network to extract discriminative features. We advocate pre-training with ALASKA2 for the model trained with BOSSBase+BOWS2. Experiments show that the proposed work outperforms previous arts in terms of detection accuracy at different payloads.

The remaining structure of this paper is organized as follows. We detail the proposed steganalysis method in Section 2, followed by extensive experimental results and analysis in Section 3. Finally, we conclude this paper and provide discussion in Section 4.

\begin{figure*}[!t]
\centering
\includegraphics[width=0.8\linewidth]{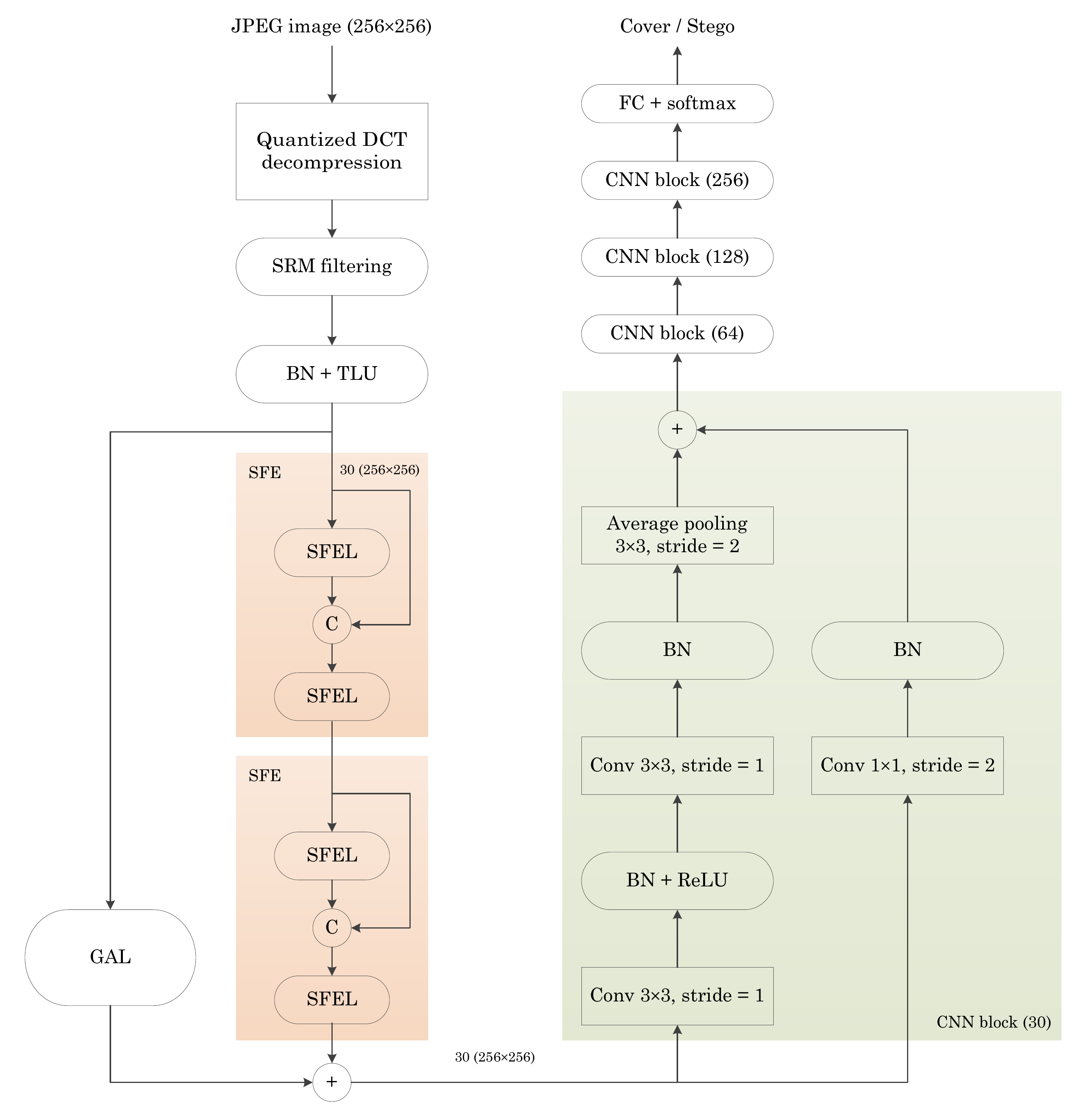}
\caption{Sketch for the proposed network architecture JPEG-GraphNet.}
\label{F1}
\end{figure*}

\section{Proposed Method}
\subsection{Overview}
The proposed network architecture is named as JPEG-GraphNet since it exploits graph representation learning for JPEG image steganalysis. Figure \ref{F1} gives an overview of JPEG-GraphNet, from which we can find that the entire process consists of three stages, i.e., feature extraction (left), feature learning (bottom-right) and feature classification (upper-right). The purpose of feature extraction is to extract the raw steganographic features and further enhance these features for facilitating deep representation learning in the stage of feature learning. The purpose of feature learning is to extract discriminative features for the stage of feature classification, whose goal is to judge whether the input is stego or not.

Specifically, feature extraction consists of three modules, i.e., preprocessing, steganographic feature enhancement (SFE) and graph attention learning (GAL). The preprocessing step extracts the steganographic residuals from the decompressed version of the input JPEG image. The noise-like residuals are then fed into the SFE module for local feature enhancement and the GAL module for global feature enhancement. The two kinds of enhanced features will be merged and fed into the subsequent feature learning module, which consists of four CNN blocks with similar structure. The feature learning module serves to represent the received steganographic features as high-dimensional features. Feature classification is the last step, where the high-dimensional features go through a fully connected layer equipped with the softmax function, and the final output is a one-hot vector demonstrating the classification result of the image. We are to detail JPEG-GraphNet in the following.

\subsection{Feature Extraction}
\subsubsection{Preprocessing}
Domain knowledge is introduced into JPEG-GraphNet through preprocessing, which packages quantized DCT decompression, SRM high-pass filtering, and dedicated activation function. Mathematically, given a JPEG image sized $h\times w$, where $h$ and $w$ are both multiples of $8$ and in our experiments they are both $256$. Let $\textbf{C} = \{c_{i, j}^{n, m}\}$ be the matrix of quantized DCT coefficients, where $c_{i, j}^{n, m}$ is the $(i,j)$-th element of the $(n,m)$-th $8\times 8$ block, $0\leq i, j< 8$, $0\leq n < h/8$, and $0\leq m < w/8$. Let $\textbf{Q} = \{q_{i, j}^{n, m}\}$ be the corresponding quantization matrix. The DCT coefficient matrix before quantization for $\textbf{C}$ can be approximated as the element-wise product of $\textbf{C}$ and $\textbf{Q}$, i.e., $\textbf{D} = \{d_{i, j}^{n, m}\}$, where $d_{i, j}^{n, m} = c_{i, j}^{n, m}q_{i, j}^{n, m}$. It is noted that generally we have $q_{i, j}^{n, m} = q_{i, j}^{n', m'}$ for any $n \neq n'$ or $m \neq m'$. To determine the spatial pixels (i.e., the decompressed luminance channel), the inverse DCT transformation is performed, i.e.,
\begin{equation}
f_{x,y}^{n,m}=\frac{1}{4}\sum_{u=0}^{7}\sum_{v=0}^{7}g_ug_vd_{u,v}^{n,m}\text{cos}\frac{(2x+1)u\pi}{16}\text{cos}\frac{(2y+1)v\pi}{16},
\end{equation}
where $g_z=1/\sqrt{2}$ for $z=0$, and $g_z = 1$ otherwise. In other words, the spatial luminance matrix of $\textbf{C}$ can be represented by $\textbf{F} = \{f_{i,j}^{n,m}\}$. The goal of quantized DCT decompression is to determine $\textbf{F}$. To determine $\textbf{D}$ from $\textbf{F}$, the DCT transformation can be applied, i.e.,
\begin{equation}
d_{u,v}^{n,m}=\frac{1}{4}g_ug_v\sum_{x=0}^{7}\sum_{y=0}^{7}f_{x,y}^{n,m}\text{cos}\frac{(2x+1)u\pi}{16}\text{cos}\frac{(2y+1)v\pi}{16}.
\end{equation}

Accordingly, $\textbf{C}$ can be determined by applying $c_{i,j}^{n,m} = d_{i, j}^{n, m}/q_{i, j}^{n, m}$. It has been widely demonstrated that high-pass filtering is essential in image steganalysis so as to capture steganographic artifacts. As a residual extractor, high-pass filtering suppresses the image content and enlarges the noise-to-signal ratio of steganalysis. High-pass filtering can be implemented with a set of trainable kernels. Although it is always open for us to initialize the trainable kernels, it is known that CNNs with low complexity usually cannot learn a good residual extractor over a relatively small dataset if they are initialized by random. We therefore recommend initializing the high-pass filtering kernels by a handcrafted way. Specifically, the SRM (which is short for Spatial Rich Model) \cite{Fridrich:TIFS:2012} filtering strategy is applied in this paper due to its superior performance. In detail, a total of $30$ basic linear filters sized $5\times 5$ described in \cite{Fridrich:TIFS:2012, Boroumand:TIFS:2019} (i.e., the ``spam'' filters and their symmetrical versions) are used to initialize the kernels, which accept $\textbf{F}$ as input and output a total of $30$ residual feature maps each sized $h\times w~(\times 1)$.

The activation function of a node in a neural network defines the output of that node given the input. The activation function can introduce nonlinearity to the neural network, which greatly increases the power of feature representation. There are various choices for the activation function such as hyperbolic tangent, sigmoid and rectified linear unit (ReLU) \cite{Nair:ICML:2010}. In conventional image steganalysis, manually crafted feature extractors typically have some kind of symmetry, e.g., histogram bins are in the intervals that are symmetric about the zero point. Though these features will be further merged to achieve high-performance steganalysis, it inspires us to introduce feature symmetry into initial feature extraction for better feature learning. Motivated by this insight, we expect that the activation function used in early stage could map the input into a symmetrical interval. Besides, empirically, it can be better to design the activation function as a hard curve, rather than a soft curve, for facilitating feature learning. Along this direction, we suggest to use the activation function called truncated linear unit (TLU) for JPEG-GraphNet, i.e., we apply TLU after SRM filtering. TLU can be expressed as \cite{Ye:TIFS:2017}:
\begin{equation}
\text{TLU}(x) = \left\{\begin{matrix}
\begin{split}
-T, &~x \in (-\infty, -T)\\
x,  &~x \in [-T, T] \\
T,  &~x \in (T, +\infty)
\end{split}
\end{matrix}\right.
\end{equation}
where $T$ is a pre-determined threshold bounding the output. In addition, to accelerate model convergence, batch normalization (BN) \cite{Ioffe:ICML:2015} is applied by default.

\begin{figure*}[!t]
\centering
\includegraphics[width=0.6\linewidth]{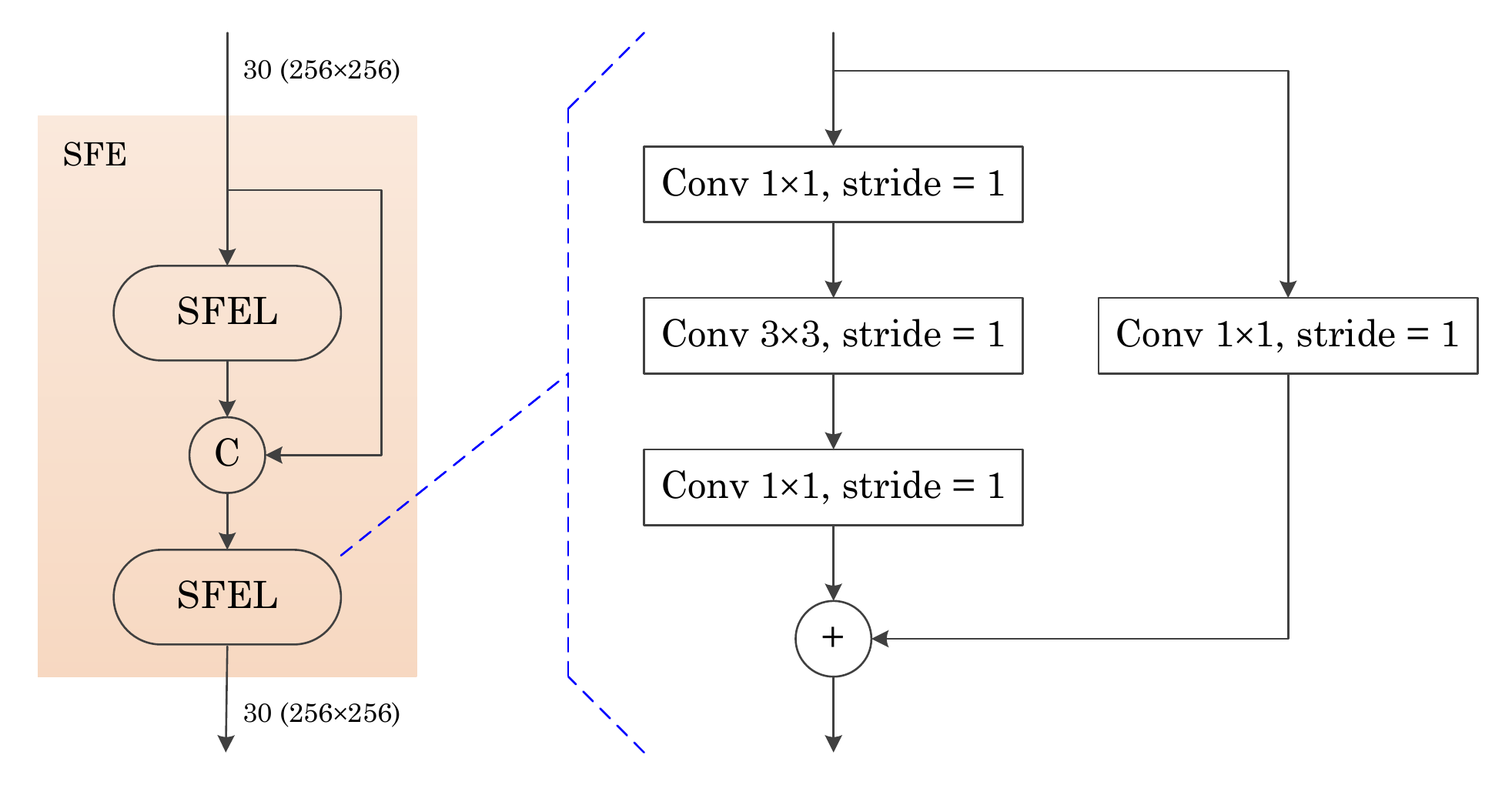}
\caption{Structural information for the SFE module.}
\label{F2}
\end{figure*}

\subsubsection{Steganographic Feature Enhancement}
To further enhance the stego information in the residual maps generated by the preprocessing step, we propose a steganographic feature enhancement (SFE) module, which is used twice as shown in Figure \ref{F1}. The detailed structural information for the SFE module is shown in Figure \ref{F2}, from which we can find that two SFE layers (SFELs) are introduced. The design principle of SFEL is inspired by the idea of improving EfficientNet for JPEG steganaysis \cite{Yousfi:IH:2021} that decreasing the resolution of the
feature maps in early CNN layers via striding or pooling negatively affects steganalysis detection accuracy as such operations suppress the noise-like stego signal while enhancing image content. So, the design of SFEL follows two key points: on one hand, pooling is removed to reduce the feature loss; on the other hand, the step of all convolutional layers are set to $1$ and the shortcut equipped with $1\times 1$ convolution is used to process the features so that the shallow feature information can  be retained as much as possible.

Since important features learned by the network at shallow layers can easily disappear at deeper layers, to prevent shallow features from disappearing, a simple way is to set up more shortcuts as the network is progressively deeper. If shallow features disappear during passing, they can still be passed directly to a deeper layer via shortcuts. We use `concat' in SFE to convert shallow features into deep features, thereby preserving more stegnaographic information.

\begin{figure*}[!t]
\centering
\includegraphics[width=0.6\linewidth]{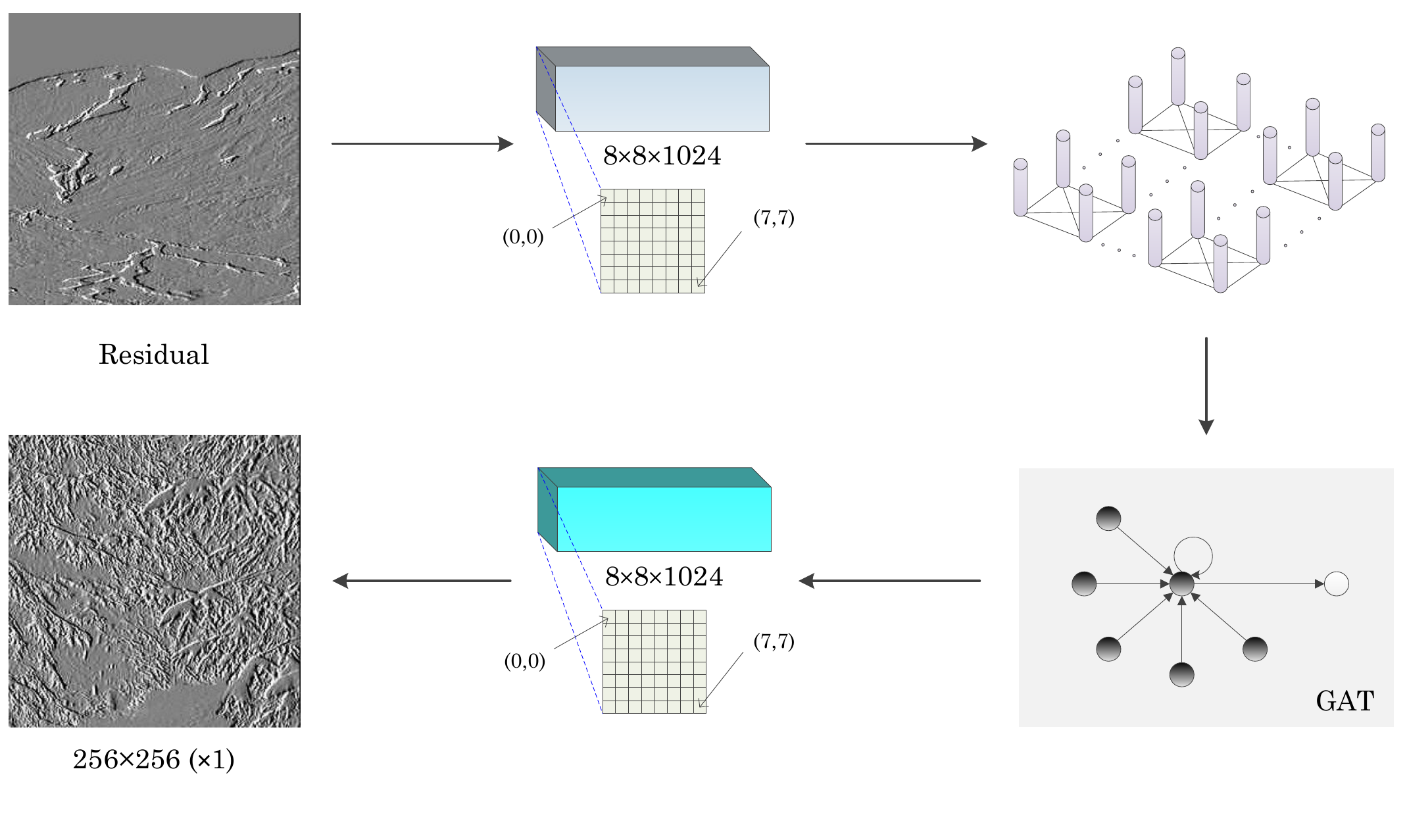}
\caption{Detailed information for GAL, where the resolution is $256\times 256$ for example.}
\label{F3}
\end{figure*}

\subsubsection{Graph Attention Learning}
To mine global steganographic information for efficient steganalysis, we introduce a graph attentional learning (GAL) module, whose details are provided in Figure \ref{F3}. In detail, given a JPEG image, it is firstly preprocessed to generate a set of residuals. Each of these residuals is then divided into disjoint $8\times 8$ blocks and superposed into a feature map with a dimension of $8\times 8\times (hw/64)$, e.g., if the width and the height of the image are both $256$, the size of the feature map is then $8\times 8\times 1024$, as described in Figure \ref{F3}. We construct a complete graph consisting of $64$ nodes, each of which is corresponding to a spatial point. Each node in the complete graph is further associated with a feature vector with a length of $hw/64$. In other words, the feature map determined above will be separated into a total of $hw/64$ feature vectors, each of which is assigned to the node corresponding to the same spatial point as the feature vector. In this way, we can build a complete graph associated with node features, where the number of nodes is $64$, the number of edges is $2016$, and the dimension of each feature vector is $hw/64$. This graph will be fed into a graph attention network for further processing. Although graph attention network deserves further study, in this paper, for simplicity, we use the graph attention network introduced in \cite{Vel:ICLR:2018} due to its superior performance.  We refer the reader to \cite{Vel:ICLR:2018} for details of the network structure. In brief summary, the network accepts a graph associated with a set of node features as input and outputs a new set of node features as the result. The number of layers of graph attention learning is set to $2$. The dimension of each new feature vector is equal to the dimension of the input feature vector. Finally, all these feature vectors are processed to constitute a new feature map sized $256\times 256$. The output will be expanded to $30$ channels by stacking new feature maps and added to the feature map generated by SFE.

\subsection{Feature Learning and Classification}
After the aforementioned steganographic feature processing, the output goes through the feature learning module. To reduce the dimensionality of the feature map, the steganographic feature learning layer employs pooling and residual shortcut with a step size of $2$. As shown in the CNN block on the right side of Figure \ref{F1}, pooling in the form of $3\times 3$ and averaging with stride $2$ are applied after two convolutional layers, and the residual shortcut is also used with a convolutional layer with a kernel size $1$ and stride $2$ to reduce the width and height of the output feature map to half of the input simultaneously. For the design of the CNN block used in this paper, we refer the reader to the feature learning part of SRNet \cite{Boroumand:TIFS:2019}, in which two types of modules, i.e., `type3' and `type4', are designed for dimensionality reduction. We design the CNN block in this paper according to type3.

A total of four CNN blocks are used for stepwise dimensionality reduction. As shown in Figure \ref{F1}, the numbers in parentheses represent the number of convolutional kernels in all convolutional layers in the block. It is worth emphasizing that the first three CNN blocks have identical structures, while the last CNN block removes the residual shortcut and replaces the average pooling at the end with a global pooling to reduce the features to one dimension. The last segment, which is a standard 256-D fully connected layer followed by a softmax linear classifier, will output final probabilities.

\section{Experimental Results and Analysis}
\subsection{Setup}
The basic training dataset used in our experiment comes from two image sources, i.e., BOSSbase v1.01 \cite{BossBase} and BOWS2 \cite{BOWS2}, each containing 10,000 grayscale images with a resolution of $512\times 512$. All the involved images are center-cropped to a resolution of $256\times 256$ due to the limited computing resource. The training set contains 14,000 cover-stego pairs (4,000 randomly chosen from BOSSbase and 10,000 from BOWS2). The validation set containing 1,000 BOSSbase cover-stego pairs is utilized to select hyper-parameters, and final testing results are reported after classification of 5,000 BOSSbase cover-stego pairs in test set. The performance of the proposed steganalysis model is evaluated at payload which ranges from 0.1 to 0.5 bits per non-zero AC DCT coefficient (bpnzac) in J-UNIWARD \cite{JUNIWARD} and UED-JC \cite{UED} for quality factors (QFs) 75 and 95.

We used PyTorch for simulation, accelerated with a single TITAN RTX 24 GB GPU. The stochastic gradient descend optimizer Adamax \cite{adam} was used with minibatches of 16 cover-stego pairs. The batch normalization parameters were learned via an exponential moving average with decay rate 0.9. In pretraining, the filter weights were initialized with the He initializer and $2\times 10^{-4}$ $L_2$ regularization. The filter biases were set to 0.2 and no regularization. The initialization of training used the pretrained network weights. For the fully connected classifier layer, we initialized the weights with a zero mean Gaussian with standard deviation of 0.01 and no bias.

\subsection{Pretraining Policy}
Pretraining can be considered as a way to initialize the network weights. The correct initialization of parameters can play an important role in training, especially in difficult tasks such as steganalysis with small payloads. In the classification problem of natural images, pre-training the model empowers the model to extract filters that form the basic elements of natural images (e.g., edges, textures, periodic patterns, etc.). Therefore, we believe that pretraining on a larger steganalysis dataset can improve the model's ability to extract steganographic information.

The pretraining dataset is the ALASKA2 dataset \cite{ALASKA2} used in the recent Kaggle competition\footnote{\url{https://www.kaggle.com/competitions/alaska2-image-steganalysis/overview}}. This dataset consists of color images compressed with three QFs 75, 90, and 95. For every QF, there are 25,000 images of resolution $512\times 512$.  We used stego images generated using J-UNIWARD and UED-JC with the payload 0.5 to 0.1 bpnzac and for QFs 75 and 95. The entire 25, 000 images pairs were all used for model pretraining.

We first trained our model on pretraining dataset for 100 epochs for different steganographic algorithms. Then, the training was executed for 350,000 iterations (400 epochs) with an initial learning rate of $r_1 = 0.001$ after which the learning rate was decreased to $r_2 = 0.0001$ for an additional 87,500 iterations (100 epochs) on the basic training dataset. The model parameter achieving the best validation accuracy in the last 100,000 iterations was taken as the result of training.

To verify how to pretrain is most effective in improving the detection accuracy of the steganalysis detector, we set up ALASKA2 pretraining datasets with three different embedding rates of 0.2, 0.4, and 0.5 bpnzac and two image qualities of QF75 and QF95.
We used the above three settings to generate J-UNIWARD based datasets and used the model parameters obtained from the pretraining of these three datasets for subsequent training with the same settings, except that the datasets were replaced with BOSSBase+BOWS2. The detection performance of the model was measured by the total classification error probability on the testing set $P_{\mathrm{E}}=\min _{P_{\mathrm{FA}}} \frac{1}{2}\left(P_{\mathrm{FA}}+P_{\mathrm{MD}}\right)$, where $P_{\mathrm{FA}}$ and $P_{\mathrm{MD}}$ are the false-alarm and missed-detection probabilities.

Experimental results are shown in Table \ref{tab1}. It can be seen that the higher the embedding rate of the pretraining set with the same embedding rate of the training set, the better the final model effect is. We speculate that the high embedding rate of the pretraining set may ``teach'' the model more a priori knowledge to help the subsequent recognition of brand new images. Therefore, we used ALASKA2 with 0.5 bpnzac as the pretraining set in all subsequent experiments.

\begin{table}[!ht]
\renewcommand{\arraystretch}{1.0}
\caption{The detection error $P_E$ for the proposed steganalysis model pretrained on ALASKA2 and trained on BOSSbase family with the J-UNIWARD algorithm.}
\centering
\resizebox{\textwidth}{!}{
\begin{tabular}{c|c|c|c}
\hline\hline
Setting & ALASKA2 0.5 bpnzac/QF75 & ALASKA2 0.4 bpnzac/QF75 & ALASKA2 0.2 bpnzac/QF75\\
\hline
BOSSbase+BOWS2 0.5 bpnzac/QF75 & .0442 & - & -\\
BOSSbase+BOWS2 0.4 bpnzac/QF75 & .0901 & .1230 & -\\
BOSSbase+BOWS2 0.2 bpnzac/QF75 & .1492 & .1679 & .1823\\
\hline\hline
\end{tabular}}\label{tab1}
\end{table}

\subsection{Performance Evaluation}
For comparison with the current state of the art methods on the union of BOSSBase and BOWS2, the proposed method was compared with SRNet \cite{Boroumand:TIFS:2019} and the network proposed by Xu \cite{Xu:IH:2017}, which we call in this paper J-XuNet. All CNN detectors compared here were trained as described in the corresponding papers. All detectors were trained on exactly the same datasets as this paper, implemented in PyTorch, and run on a single GPU. As illustrated in Table \ref{tab2}, with multiple embedding rates and multiple embedding algorithms, the proposed method has the best detection accuracy performance.

\begin{table*}[!ht]
\renewcommand{\arraystretch}{1.0}
\centering
\caption{The detection error $P_E$ for the proposed method and previous arts for five payloads in bpnzac for J-UNIWARD and UED-JC for QFs 75 (left) and 95 (right).}
\resizebox{\textwidth}{!}{
\begin{tabular}{c|c|ccccc|ccccc}
\hline\hline
\multicolumn{1}{c|}{\multirow{2}{*}{Embedding method}} & \multicolumn{1}{c|}{\multirow{2}{*}{Model}} &
\multicolumn{5}{c|}{QF75} &
\multicolumn{5}{c}{QF95}
\\ \cline{3-12}
& & 0.5 & 0.4 & 0.3  & 0.2 & 0.1 & 0.5 & 0.4 & 0.3  & 0.2 & 0.1\\
\hline
& J-XuNet & .0776 & .1207 & .1895 & .2892 & .4310 & .2543 & .3162 & .4246 & .4512 & .4812\\
\multirow{2}{*}{J-UNIWARD} & SRNet & .0575 & .1070 & .1235 & .1929 & .3621 & .1268 & .1845 & .2831 & .3774 & .4367\\
& Proposed & .0442 & .0901 & .1049 & .1492 & .3014 & .1083 & .1595 & .2387 & .3050 & .4021 \\
\hline
& J-XuNet& .0473 & .0887 & .1108	& .1372	& .2944	& .0983	& .1592	& .2091	& .2983	& .3958 \\
UED-JC & SRNet & .0113 & .0789 & .0985 & .1188 & .1923 & .0700 & .0978 & .1362 & .2054	& .3369 \\
& Proposed & .0105 & .0521 & .0728 & .1003 & .1739 & .0623 & .0804 & .1122 & .1815 & .2950 \\
\hline\hline
\end{tabular}}\label{tab2}
\end{table*}

The accuracy curves during the training of the model in Fig. 4 show that the proposed model has better accuracy performance and higher convergence accuracy during the same training period. From the accuracy curves, we can see that the proposed model at the beginning of training can have higher accuracy than the other models, which indicates that pretraining does give the model prior knowledge and helps the model converge faster. And in the subsequent training, the test accuracy is always higher than others, which also verifies that our proposed SFE and GAL modules can effectively enhance the extraction of steganographic information. The efficient feature-enhancing pattern of SFE combined with the graph attention learning of GAL allows the network to better pass on critical weakly hidden signals, thus allowing the proposed model to maintain a high level of accuracy.

\begin{figure*}[!ht]
\centering
\includegraphics[width=0.9\linewidth]{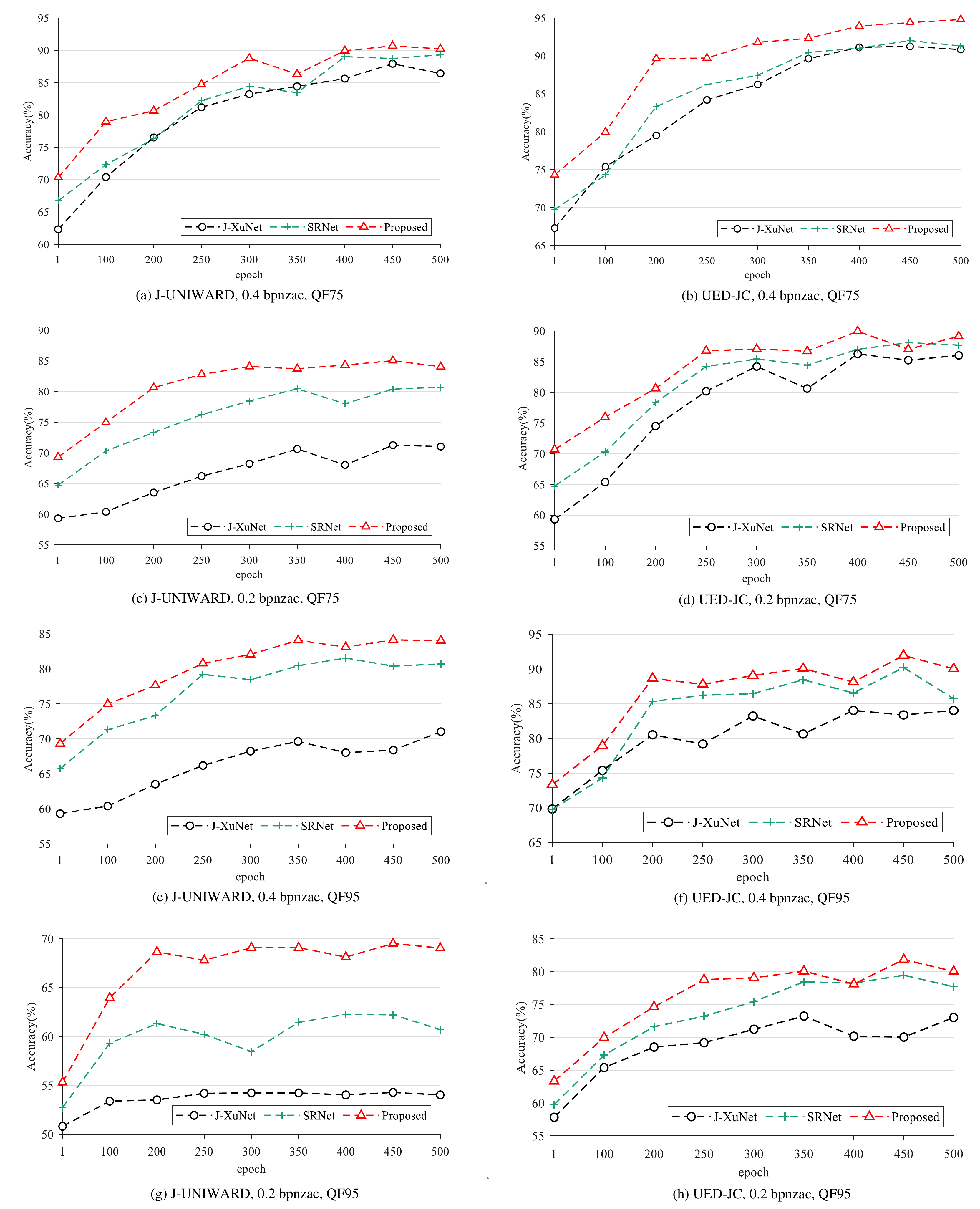}
\caption{The detection accuracy curves on the validation set.}
\label{F4}
\end{figure*}

\subsection{Ablation Study}
To validate the effectiveness of the proposed SFE module and GAL module, we have further conducted ablation experiments. We conducted ablation experiments on the more representative embedding rate of 0.4 bpnzac, and Table \ref{tab3} shows the detection error rate without SFE module, without GAL module, and without both compared to the standard model. It can be seen from the data in the table that without SFE module and without GAL module both cause an increase in the detection error rate of the model, but the accuracy of the model without SFE is sharply reduced by about 10$\%$ more than that without GAL so SFE has a stronger importance in steganographic information enhancement, while GAL may be able to assist SFE in further improve performance through its ability to extract global information signal. The sharp decrease in detection accuracy of the model with simultaneous removal of SFE and GAL indicates that they are both very important in the part of steganographic information extraction, and validates the usefulness of our proposed two modules.

\begin{table}[!ht]
\renewcommand{\arraystretch}{1.0}
\caption{The detection error $P_E$  without the SFE module, without the GAL module, and without both compared to the standard model.}
\centering
\resizebox{\textwidth}{!}{
\begin{tabular}{c|c|c|c|c}
\hline\hline
& J-UNIWARD QF75 & J-UNIWARD QF95 & UED-JC QF75 & UED-JC QF95 \\
\hline
Proposed & .0901 & .1595 & .0521 &.0804\\
without SFE & .0232 & .3076 & .0203 & .2897\\
without GAL & .0133 & .1830  & .0109 &.1433\\
without SFE and GAL & .3987 & .4483 & .3024 & .4075\\
\hline\hline
\end{tabular}}\label{tab3}
\end{table}

\section{Conclusion}
In this paper, we combine the characteristics of weak stego signal transfer in CNNs and the advantages of globalization of graph attention neural networks to design a method that is more suitable for feature transfer and enhancement of image steganographic signals. In general, compared with other methods, the modules proposed in this paper can better solve the problem of weakening steganographic features due to the stacking of network layers. In addition, it is also capable of making better use of the global features of the steganographic signal. The model obtained after our pretraining with a larger dataset is also able to achieve better detection accuracy compared with existing excellent methods. With the popularity of graph neural networks, many unsupervised works have emerged. In the future, we will explore methods to apply unsupervised learning to image steganography analysis.

\section*{Acknowledgement}
This work was supported by the CCF-Tencent Rhino-Bird Young Faculty Open Research Fund.

\end{document}